\documentclass[a4paper, oneside, twocolumn, notitlepage, 10pt]{extarticle_ecoc}
\usepackage{ecoc}

\begin{document}
\selectlanguage{english}    


\title{On the Feasibility of SCL-Band Transmission over G.654.E-Compliant Long-Haul Fibre Links}%


\author{
    Jiaqian~Yang\textsuperscript{(1)}, Eric~Sillekens\textsuperscript{(1)}, Ronit~Sohanpal\textsuperscript{(1)}, Mingming~Tan\textsuperscript{(2)}, Dini~Pratiwi\textsuperscript{(2)}, \\Henrique~Buglia\textsuperscript{(3)}, Romulo~Aparecido\textsuperscript{(1)}, John~D.~Downie\textsuperscript{(4)}, Sergejs~Makovejs\textsuperscript{(4)}, Lidia~Galdino\textsuperscript{(4)},
    \\Wladek~Forysiak\textsuperscript{(2,5)}, Polina~Bayvel\textsuperscript{(1)}, Robert~I.~Killey\textsuperscript{(1)}
}

\maketitle                  


\begin{strip}
    \begin{author_descr}
    
        \textsuperscript{(1)} Optical Networks Group, UCL (University College London), London, UK, 
        \textcolor{blue}{\uline{jiaqian.yang.18@ucl.ac.uk}}\\
        \textsuperscript{(2)} Aston Institute of Photonic Technologies, Aston University, Birmingham, B4 7ET, UK\\
        \textsuperscript{(3)} Nokia, San Jose, CA, USA\\
        \textsuperscript{(4)} Corning Incorporated, 1 Riverfront Plaza, NY 14831, USA\\
        \textsuperscript{(5)} High Performance Networks Group, University of Bristol, Bristol, BS8 1UB, UK\\
    \end{author_descr}
\end{strip}

\renewcommand\footnotemark{}
\renewcommand\footnoterule{}


\begin{strip}
    \begin{ecoc_abstract}
        We demonstrate the first SCL-band long-haul transmission using G.654.E-compliant fibre, achieving 100.8\,Tb/s (GMI) over 1552\,km, despite its 1520\,nm cutoff wavelength. Due to the fibre's ultra-low loss and low nonlinearity, the achievable-information-rate with lumped amplification is comparable to that of G.652.D-compliant fibre links with distributed-Raman-amplification. \textcopyright2025 The Author(s)
        \vspace{-1em}
    \end{ecoc_abstract}
\end{strip}


\section{Introduction}
\begin{figure}
    \centering
    \begin{tikzpicture}\footnotesize
    \begin{groupplot}[
    group style={group size=1 by 2,vertical sep=0pt},
    xmin=800, xmax=2500,
    xmode=log,
    xtick={800,900,1000,1100,1200,1300,1400,1500,1600,1700,1800,1900,2000,2100,2200,2300,2400,2500},
    width=\linewidth,
    height=4cm,
    xticklabel style={/pgf/number format/1000 sep=},
    y label style={at={(axis description cs:0.08,0.5)}},
    ]
    \newcommand{\datapoint}[5]{
    \addplot[color=#4,only marks, mark=#5*, mark size=2pt,
    point meta=explicit symbolic, 
    nodes near coords={\cite{#1}},
    nodes near coords style={inner sep=1pt, anchor=#3},forget plot,
    ] coordinates {#2};
    }
    \newcommand{\datapointt}[5]{
    \addplot[color=#4,only marks, mark=#5*, mark size=2.5pt,
    point meta=explicit symbolic, 
    nodes near coords={\cite{#1}},
    nodes near coords style={inner sep=1pt, anchor=#3},forget plot,
    ] coordinates {#2};
    }

    \nextgroupplot[
    ylabel={Throughput (Tb/s)},
    xmode=log,
    ymax=170,
    ymin=50,
    xticklabels=\empty,
    ytick={70,100,130,160},
    yticklabels={70,100,130,160},
    legend style={column sep=2pt, fill opacity=1, text opacity = 1, at={(axis cs: 2500,170)}},
    legend cell align={left},
    ]

    \datapoint{hamaoka2024_110}{(1040,110.7)}{south west}{j2}{};
    \datapointt{kobayashi2024c}{(800,115.3)}{north west}{j4}{triangle};
    \datapointt{kobayashi2024c}{(2400,72.6)}{north east}{j4}{triangle};
    \datapoint{shimizu2024_133}{(1040,133.06)}{south west}{q3}{square};
    \datapoint{yang2025transmission}{(1014,67.3)}{south west}{j2}{};
    \datapoint{hamaoka2025_107}{(1200,107.7)}{north west}{j2}{};
    \datapointt{shimizu2025_27}{(1040,160.2)}{north east}{q4}{diamond};
    
    \addlegendentry{SCL}
    \addlegendimage{j2,only marks, mark=*,mark size=2pt, line width=1pt,}
    \addlegendentry{CLU}
    \addlegendimage{j4,only marks, mark=triangle*,mark size=2.5pt, line width=1pt,}
    \addlegendentry{SCLU}
    \addlegendimage{q3,only marks, mark=square*,mark size=2pt, line width=1pt,}
    \addlegendentry{SCLUX}
    \addlegendimage{q4,only marks, mark=diamond*,mark size=2.5pt, line width=1pt,}


    \addplot[Set1-A, only marks,mark=o, mark options={mark size = 2}, line width=1pt,
    point meta=explicit symbolic, 
    nodes near coords={SCL},
    nodes near coords style={
        inner sep=2pt,
        anchor=south west,
    },
    ] coordinates {(1552,100.8)};
    \addplot[Set1-A, only marks,mark=*, mark options={mark size = 2},
    point meta=explicit symbolic, 
    nodes near coords style={
        inner sep=2pt,
        anchor=north west,
    },
    ] coordinates {(1552,92.8)};

    \addplot[dashed, gray, domain=800:2000, samples=100, forget plot]
    {100000/x};
    \addplot[dashed, gray, domain=900:2500, samples=100, forget plot]
    {150000/x};
    \addplot[dashed, gray, domain=1200:2500, samples=100, forget plot]
    {200000/x};
    \node[gray,rotate=-12,opacity=0.8,font=\fontsize{7}{0}\selectfont] at (axis cs: 1300,70) {100 Pb/s$\cdot$km};
    \node[gray,rotate=-22,opacity=0.8,font=\fontsize{7}{0}\selectfont] at (axis cs: 1280,128) {150 Pb/s$\cdot$km};
    \node[gray,rotate=-25,opacity=0.8,font=\fontsize{7}{0}\selectfont] at (axis cs: 1500,142) {200 Pb/s$\cdot$km};

    \nextgroupplot[
    height=3.5cm,
    xlabel={Distance (km)},
    ylabel={SE (b/s/Hz)},
    xmode=log,
    ymax=8,
    ymin=3.5,
    xticklabels={800,,1000,,,,,1500,,,,,2000,,,,2400,},
    ytick={4,5,6,7},
    yticklabels={4,5,6,7},
    clip=false,
    x label style={at={(axis description cs:.5,0.15)}},
    ]
    \renewcommand{\datapoint}[5]{
    \addplot[color=#4,only marks, mark=#5*, mark size=2pt,
    point meta=explicit symbolic, 
    nodes near coords={\cite{#1}},
    nodes near coords style={inner sep=1pt, anchor=#3},forget plot,
    ] coordinates {#2};
    }\renewcommand{\datapointt}[5]{
    \addplot[color=#4,only marks, mark=#5*, mark size=2.5pt,
    point meta=explicit symbolic, 
    nodes near coords={\cite{#1}},
    nodes near coords style={inner sep=1pt, anchor=#3},forget plot,
    ] coordinates {#2};
    }
    
    \datapointt{kobayashi2024c}{(800,7.76)}{north west}{j4}{triangle};
    \datapointt{kobayashi2024c}{(2400,4.89)}{north east}{j4}{triangle};
    \datapoint{shimizu2024_133}{(1040,6.04)}{north west}{q3}{square};
    \datapoint{yang2025transmission}{(1014,4.83)}{north west}{j2}{};
    \datapoint{hamaoka2025_107}{(1200,5.89)}{north west}{j2}{};
    \datapoint{hamaoka2024_110}{(1040,6.049)}{south west}{j2}{};
    \datapointt{shimizu2025_27}{(1040,5.93)}{north east}{q4}{diamond};
    
    \addplot[Set1-A, only marks,mark=*, mark options={mark size = 2},line width=1pt,
    point meta=explicit symbolic, 
    nodes near coords style={
        inner sep=2pt,
        anchor=north west,
    },
    ] coordinates {(1552,6.15)};
    \addplot[Set1-A, only marks,mark=o, mark options={mark size = 2},line width=1pt,
    point meta=explicit symbolic, 
    nodes near coords style={
        inner sep=2pt,
        anchor=north west,
    },
    ] coordinates {(1552,6.68)};

    \draw[>=stealth, ->](axis cs:1900,7)--(axis cs:1600,9.5);
    \draw[>=stealth, ->](axis cs:1900,6.5)--(axis cs:1600,6.2);
    \node[anchor=west] at (axis cs:1900,6.75) {This work};

    \end{groupplot}
    \end{tikzpicture}
    \caption{Transmission distance, throughput, and spectral efficiency (SE) of recent long-haul transmission demonstrations using three or more optical bands. Solid markers: decoded net rate; open markers: rate from GMI.}
    \label{fig:state_of_the_art}
\end{figure}
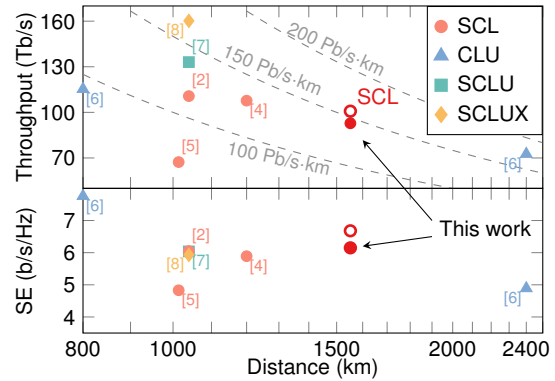

\vspace{-.5em}
In recent years, major breakthroughs have been achieved in ultra-wideband (UWB) wavelength-division multiplexing (WDM) fibre transmission employing three or more optical bands, as summarised in Fig.~\ref{fig:state_of_the_art}. 
However, the use of UWB technology in long-haul transmission remains challenging, as separate amplifiers are required for each band, and signal power profile shaping and optimisation become increasingly complex, due to inter-channel stimulated Raman scattering (ISRS) and the use of Raman amplification. Distributed Raman amplification, as an approach to complement lumped doped fibre amplifiers (DFAs), has been widely used in long-haul UWB transmission~\cite{puttnam2022investigation,hamaoka2024_110,yang2024experimental,hamaoka2025_107} to compensate for fibre loss and ISRS. Other studies have employed DFA-only approaches~\cite{yang2025transmission}, or utilised wavelength conversion techniques to amplify signals in the shorter or longer wavelength bands~\cite{kobayashi2024c,shimizu2024_133,shimizu2025_27}. These complex amplification schemes become essential, as most transmission experiments have been carried out over G.652.D-compliant standard single-mode fibre, which has an effective area of approximately 80~\textmu m\textsuperscript{2} and attenuation of 0.2\,dB/km at 1550\,nm. The low water-absorption peak of such fibre allows Raman pump lasers operating in the E-band to be deployed, providing distributed gain to offset power loss due to ISRS. However, new optical fibres have been developed, such as G.654.E-compliant fibres, featuring ultra-low-loss characteristics, with reduced attenuation of 0.148\,dB/km at 1550\,nm and a larger effective area (125~\textmu m\textsuperscript{2}), though with a maximum cutoff wavelength shifted to $\sim$1520\,nm. These fibres are well-suited for high-data-rate, long-haul transmission~\cite{downie2018g}, and have been demonstrated as suitable for S-band transmission, both theoretically~\cite{downie2023on} and experimentally~\cite{aparecido2024experimental}, even when signal wavelengths fall below the cutoff. Although its lower nonlinearity (and hence lower Raman gain coefficient) and the presence of a higher water absorption peak may reduce the efficiency of Raman amplification in G.654.E fibre, studies have shown that Raman pumping can still provide performance gains~\cite{mlejnek2019examination,ivanov2025performance}. Several prior works using G.654.E fibre have been reported~\cite{kobayashi2024c,zhang2024_201,zhang2025_214}, though they feature either limited transmission distances of no more than 150\,km or signals at wavelengths above the cutoff wavelength.

In this work, we investigate the feasibility of SCL-band long-haul transmission over G.654.E-compliant Corning\textsuperscript{\textregistered} Vascade\textsuperscript{\textregistered} EX2500 fibre and achieve the highest SE in SCL-band long-haul transmission. Negligible penalty due to multipath interference (MPI) was observed for signals below the cutoff wavelength. With the benefit of lower attenuation, 
the G.654.E fibre was found to enable similar throughput to G.652.D fibre-based transmission, but without the need for distributed Raman amplification. Furthermore, despite the higher water absorption peak of the G.654.E fibre, the use of distributed Raman amplification with pump wavelengths in the E-band, provided improved transmission performance. The results show the potential of this type of fibre for long-haul UWB optical transmission.

\section{Experimental setup}
\input{figure/exp}
The experiment carried out (shown in Fig.~\ref{fig:exp}) consisted of an external-cavity laser (ECL) and a dual-polarisation in-phase quadrature (DP-IQ) thin film lithium niobate (TFLN) modulator with a 3-dB bandwidth of $\sim$80\,GHz, driven by an arbitrary waveform generator (AWG) to generate a 112~GBd geometrically-shaped (GS)~\cite{sillekens2022high} 16-QAM or 64-QAM test channel. A fourth-order Volterra digital pre-distortion filter was applied to compensate for the nonlinear response of the digital-to-analogue converter (DAC), driver amplifier and modulator, achieving a maximum back-to-back signal-to-noise ratio (SNR) of $\sim$19\,dB across all three bands. Co-propagating WDM channels were emulated using spectrally-shaped amplified spontaneous emission (ASE) noise, generated by a wideband ASE source and shaped by WaveShapers\textsuperscript{\textregistered} operating as wavelength selective switches (WSSs). The combined WDM signal was transmitted through a recirculating fibre loop for 18 re-circulations, totalling a distance of 1552\,km. The loop comprised a pair of acousto-optic modulators (AOMs) for switching between loading and re-circulating states, a polarisation scrambler (PS), three gain blocks (each of which included an S-band thulium-doped fibre amplifier (TDFA), a low-gain C-band EDFA and a low-gain L-band EDFA), a 86.2\,km G.654.E-compliant fibre span, three backward Raman pumps, and WSSs and variable optical attenuators (VOAs) for power balancing. The UWB receiver consisted of a bandpass filter, a pre-amplifier, a 90-degree hybrid, and a 256~GSa/s Keysight UXR oscilloscope. Offline digital signal processing (DSP)~\cite{wakayama2021_2048} and adaptive rate decoding~\cite{geiger2023performance,yang2025transmission} was performed.

\section{Fibre and Raman pump characterisation}
\begin{figure}
    \centering
    \begin{tikzpicture}
    \node[inner sep=0pt,font=\fontsize{7}{0}\selectfont] (background) at (5.1,0.5) {
        \setlength\extrarowheight{10pt}
        \setlength{\tabcolsep}{2pt}
        \begin{tabular}{c|c|c}
             & G.654.E & G.652.D \\
            1365 & 0.28 & 0.27 \\
            1385 & 0.76 & 0.27 \\
            1405 & 0.39 & 0.25 \\
            1425 & 0.25 & 0.24 \\
        \end{tabular}};
    \node[inner sep=0pt,font=\fontsize{7}{0}\selectfont] (background) at (5.2,1.7) {At pump wavelengths};
    \begin{scope}
        \begin{axis}[
        height=3.5cm,
        width=0.7\linewidth,
        xmin=1470,xmax=1620,
        ymin=0.14,ymax=0.28,
        xtick={1480,1520,1560,1600},
        xticklabels={1480,1520,1560,1600},
        ytick={0.14,0.18,0.22,0.26},
        yticklabels={0.14,0.18,0.22,0.26},
        xlabel={Wavelength (nm)},
        ylabel={Att. (dB/km)},
        xlabel near ticks,
        ylabel near ticks,
        ymajorgrids=true,
        x label style={font=\fontsize{7}{0}\selectfont},
        y label style={font=\fontsize{7}{0}\selectfont,at={(axis description cs:-0.15,0.5)}},
        x tick label style={font=\fontsize{7}{0}\selectfont},
        y tick label style={font=\fontsize{7}{0}\selectfont},
        legend style={at={(axis cs:1620,0.28)}, anchor=north east,font=\fontsize{7}{0}\selectfont},
        ]
        \addlegendentry{G.654.E};
        \addlegendimage{j4, line width=1.5pt};
        \addlegendentry{G.652.D};
        \addlegendimage{j2, line width=1.5pt};
        \addplot[j4, line width=1.5pt, forget plot] table[x=wavelength,y=loss]{data/loss_vascadeex2500.txt};
        \addplot[j2, line width=1.5pt, forget plot] table[x=wavelength,y=loss]{data/loss_sterlite.txt};
        \draw[>=stealth, ->](axis cs:1540,0.16)--(axis cs:1550,0.15);
        \node[font=\fontsize{7}{0}\selectfont, anchor=south east] at (axis cs:1560,0.15) {\setlength\extrarowheight{7pt}
        \begin{tabular}{c}
             0.148\,dB/km \\
             @1550\,nm
        \end{tabular} };
        \end{axis}
        \end{scope}

    \end{tikzpicture}
    \caption{Fibre attenuation at signal and Raman pump wavelengths. }
    \label{fig:fibre}
\end{figure}
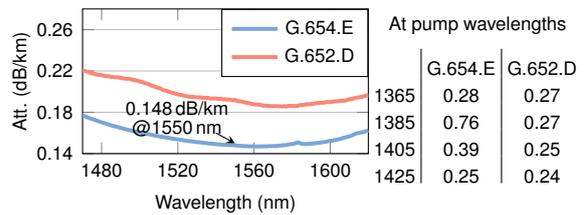
The Vascade EX2500 fibre had a span length of 86.2\,km, and a low-water-peak G.652.D fibre of the same span length was used for comparison. Throughout the remainder of this paper, we refer to the Vascade EX2500 fibre and the low-water-peak fibre as G.654.E and G.652.D, respectively. The measured attenuation of the two fibres is shown in Fig.~\ref{fig:fibre}, with G.654.E exhibiting $\sim$0.04\,dB/km lower attenuation across the signal wavelengths, but higher loss at the pump wavelengths in the E-band due to the presence of the water absorption peak. The total fibre span losses at 1550\,nm, including splicing losses, were 13.0\,dB for G.654.E and 16.5\,dB for G.652.D.

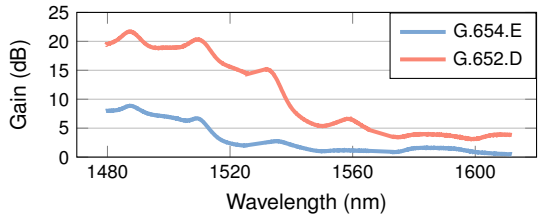
\begin{figure}
    \centering
    \begin{tikzpicture}\footnotesize
        \begin{axis}[
        height=3.5cm,
        width=\linewidth,
        xmin=1470,xmax=1620,
        ymin=0,ymax=25,
        xtick={1480,1520,1560,1600},
        xticklabels={1480,1520,1560,1600},
        ytick={0,5,10,15,20,25},
        yticklabels={0,5,10,15,20,25},
        xlabel={Wavelength (nm)},
        ylabel={Gain (dB)},
        xlabel near ticks,
        ymajorgrids=true,
        y label style={at={(axis description cs:.08,.5)}},
        x tick label style={font=\fontsize{7}{0}\selectfont},
        y tick label style={font=\fontsize{7}{0}\selectfont},
        legend style={at={(axis cs:1620,25)}, anchor=north east,font=\fontsize{7}{0}\selectfont},
        ]
        \addlegendentry{G.654.E};
        \addlegendimage{j4, line width=1.5pt};
        \addlegendentry{G.652.D};
        \addlegendimage{j2, line width=1.5pt};
        \addplot[j4, line width=1.5pt, forget plot] table[x=Wavelength,y=RGain]{data/RGain.txt};
        \addplot[j2, line width=1.5pt, forget plot] table[x=Wavelength,y=RGainSterlite]{data/RGain.txt};
        \end{axis}
    \end{tikzpicture}
    \caption{Raman on-off gain. }
    \label{fig:raman}
\end{figure}

Three lasers were used as Raman pumps in the G.654.E fibre, at wavelengths of 1365\,nm, 1385\,nm, 1405\,nm with maximum output powers of $\sim$500\,mW each. For the G.652.D fibre, an additional pump at 1425\,nm was employed to enhance the Raman gain. The pump powers and signal launch power per band were optimised for both fibre types using a simplified ISRS GN model~\cite{buglia2024closed} to maximise throughput. Wavelength-dependent gains or losses were replaced with average values per band in the model to simplify optimisation and system design. Pump powers were set to maximum for G.654.E fibre due to reduced Raman efficiency~\cite{ivanov2025performance}, while 447\,mW, 501\,mW, 331\,mW, 224\,mW powers were used for the four pumps for transmission over the G.652.D fibre, with similar total pump power of $\sim$1.5\,W. 
The optimised total launch powers of the signals in the S-, C-, and L-bands were 19.7\,dBm, 19.4\,dBm, 18.5\,dBm, respectively, for the G.654.E fibre, and 19.7\,dBm, 16.6\,dBm, 18.5\,dBm, respectively, for the G.652.D fibre. The Raman on-off gain was measured using an optical spectrum analyser (OSA), as shown in Fig.~\ref{fig:raman}. Despite using the maximum available Raman pump powers for the G.654.E fibre, the gain in the S-band remained below 10\,dB due to its lower Raman gain coefficient - approximately half that of the G.652.D fibre~\cite{ivanov2025performance}. In contrast, pumping the G.652.D fibre achieved a maximum Raman gain of more than 20\,dB.

\section{Results}
\begin{figure}[h]
    \centering
    \begin{tikzpicture}\footnotesize
        \begin{axis}[
        height=4cm,
        width=\linewidth,
        xmin=1470,xmax=1620,
        ymin=-60,ymax=0,
        xtick={1480,1520,1560,1600},
        xticklabels={1480,1520,1560,1600},
        ytick={-60,-50,-40,-30,-20,-10,0},
        yticklabels=\empty,
        xlabel={Wavelength (nm)},
        ylabel={10\,dB/div.},
        xlabel near ticks,
        ymajorgrids=true,
        y label style={at={(axis description cs:.15,.5)}},
        x tick label style={font=\fontsize{7}{0}\selectfont},
        y tick label style={font=\fontsize{7}{0}\selectfont},
        legend style={at={(axis cs:1545,-12)}, anchor=south,font=\fontsize{7}{0}\selectfont},
        legend cell align={left},
        legend columns=2,
        ]
        \addlegendentry{Shaped ASE};
        \addlegendimage{j2, line width=1pt,dashed};
        \addlegendentry{Fibre out with Raman};
        \addlegendimage{q4, line width=1pt};
        \addlegendentry{Fibre in};
        \addlegendimage{j4, line width=1pt};
        \addlegendentry{After 18 spans};
        \addlegendimage{q3, line width=1pt,};
        \addplot[j4, line width=1pt, forget plot] table[x=Wavelength,y=VascadeFin]{data/Spectra.txt};
        \addplot[q4, line width=1pt, forget plot] table[x=Wavelength,y=VascadeFout]{data/Spectra.txt};
        \addplot[q3, line width=1pt, forget plot] table[x=Wavelength,y=Power_18]{data/ASE.txt};
        \addplot[j2, line width=1pt, dashed,forget plot] table[x=Wavelength,y=Power]{data/ASE.txt};
        \end{axis}
    \end{tikzpicture}
    \caption{Spectra at loop monitor, G.654.E fibre input and output.}
    \label{fig:spec}
    \vspace{.5em}
\end{figure}
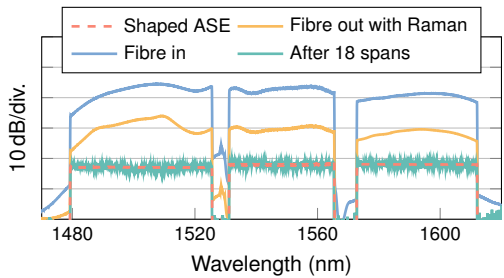

The signal spectral power was monitored using an OSA and is shown in Fig.~\ref{fig:spec}. 
The wavelength-dependent gain of the lumped amplifiers led to a non-flat launch power profile into the fibre, which was the primary cause of the SNR variations.

After DSP and deducting pilot overhead, the SNR and GMI for the G.654.E fibre transmission - with and without Raman amplification - are shown in Fig.~\ref{fig:snr}(a) and (b). In the S-band, the SNR ranges from 5\,dB to 11\,dB, where a geometrically-shaped (GS)-16\,QAM constellation was employed. In contrast, the received signals in the C- and L-bands exhibited higher SNRs of $\sim$12\,dB, allowing the use of a higher-cardinality GS-64\,QAM constellation. Comparing Fig.~\ref{fig:snr}(a) and (b), the most significant improvement is observed in the shorter-wavelengths of the S-band, where an average GMI increase of 0.86\,bit/4D-symbol is achieved for channels between 1480\,nm and 1500\,nm with Raman amplification. The remaining channels in the S-band and those in the C- and L-bands show similar performance. The total throughput over the 1552\,km G.654.E fibre, estimated from GMI, is 100.85\,Tb/s with Raman amplification and 97.02\,Tb/s without. Compared to the results of the G.652.D fibre transmission experiment with Raman amplification, shown in Fig.~\ref{fig:snr}(c), similar SNR is achieved in the S-band, indicating negligible MPI-induced penalties in transmission over G.654.E fibre at wavelengths below the cutoff wavelength. In the C- and L-bands, approximately 0.67 and 0.20\,bit/4D-symbol higher GMI values, respectively, were achieved with the G.654.E fibre, largely due to the significantly lower attenuation and larger effective area of the G.654.E fibre. 

After decoding, the net data rates in Fig.~\ref{fig:snr}(d), corresponding to the scenarios in (a)-(c), were 92.8\,Tb/s and 88.8\,Tb/s for G.654.E fibre transmission with and without Raman, respectively. The G.654.E fibre with only lumped DFA achieves a throughput comparable to that of the G.652.D fibre with Raman amplification (89.9\,Tb/s), which requires over 1.5\,W of pump power. These results demonstrate the suitability of G.654.E-compliant fibre for long-haul UWB transmission.

\input{figure/snr}

\section{Conclusion}
We successfully transmitted SCL-band 15.08\,THz bandwidth signals at a high baud rate of 112\,GBd over 1552\,km of cutoff-shifted G.654.E Vascade EX2500 fibre, achieving more than 100\,Tb/s throughput with negligible performance penalty observed below the cutoff wavelength. Transmission performance comparable to that of a Raman-amplified low-water-peak G.652.D fibre link highlights the potential of long-haul UWB transmission using DFA only with G.654.E fibre.

\section{Acknowledgements}
This work was supported by EPSRC Grant EP/R035342/1 Transforming Networks - building an intelligent optical infrastructure (TRANSNET), EP/W015714/1 Extremely Wideband Optical Fibre Communication Systems (EWOC), and EP/V007734/1 EPSRC Strategic equipment grant.
 
The authors would like to thank Yuta Wakayama (KDDI Research) for loan of the S-band amplifiers and Ralf Stolte (Coherent / Finisar) for loan of the WaveShaper. The authors would like to acknowledge Sumitomo Osaka Cement Co., Ltd. for providing the high-bandwidth Coherent Driver Modulator, and PlugTech for the automatic bias controller.

\printbibliography
\vspace{-4mm}

\end{document}